\documentclass[aps,prl,reprint,twocolumn,superscriptaddress,amsmath,amssymb,citeautoscript,notitlepage]{revtex4-1}
\usepackage{hyperref}
\hypersetup{colorlinks=true, citecolor=blue, urlcolor=blue, linkcolor=blue}
\usepackage{amsmath}
\usepackage{amssymb}
\usepackage{graphicx}
\usepackage{float}
\usepackage[section]{placeins}
\usepackage{color}

\newcommand{\munich}{Department of Physics, Arnold Sommerfeld Center for Theoretical Physics (ASC), Munich Center for Quantum Science and Technology (MCQST), Fakult\"at f\"ur Physik, Ludwig-Maximilians-Universit\"at M\"unchen, D-80333 M\"unchen, Germany} 
\newcommand{\geneve}{Department of Quantum Matter Physics, University of Geneva, 1211 Geneva, Switzerland}

\begin{document}
\title{Probing the Hall Voltage in Synthetic Quantum Systems}
\author{Maximilian Buser}
\affiliation{\munich} 
\author{Sebastian Greschner}
\affiliation{\geneve} 
\author{Ulrich Schollw\"ock}
\affiliation{\munich} 
\author{Thierry Giamarchi}
\affiliation{\geneve} 
\begin{abstract}
In the context of experimental advances in the realization of artificial magnetic fields in quantum gases, we discuss feasible schemes to extend measurements of the Hall polarization to a study of the Hall voltage, allowing for direct comparison with solid state systems.
Specifically, for the paradigmatic example of interacting flux ladders, we report on characteristic zero crossings and a remarkable robustness of the Hall voltage with respect to interaction strengths, particle fillings, and ladder geometries, which is unobservable in the Hall polarization.
Moreover, we investigate the site-resolved Hall response in spatially inhomogeneous quantum phases. 
\end{abstract}
\date{\today}
\maketitle
In the age of synthetic quantum systems, the realization of artificial gauge fields in ultracold gases~\cite{lin_09a, lin_09b, aidelsburger_11, miyake_13, atala_13, atala_14, celi_14, livi_16, kolkowitz_17, an_17,tai_17, an_18} opens up an exciting path for the study of interacting particles in the presence of large magnetic fluxes.
In these platforms, the Hall-like response of a particle current constitutes a typical fingerprint of the presence of an emulated magnetic field:
pioneering experiments measured the transverse polarization $p_y$ in synthetic few-leg flux ladders after inducing a transient longitudinal current $j_x$~\cite{aidelsburger_13, stuhl_15, mancini_15, genkina_19, chalopin_20}, readily giving rise to the Hall polarization $P_\mathrm{H}=p_y/j_x$.
%

%
Above and beyond that, there is the prospect of quantum gases probing the Hall response in the strongly interacting regime.
As theoretical calculations remain challenging therein~\cite{brinkman_71, shastry_93, prelovsek_99, zotos_00, lopatin_01, leon_07, huber_11, auerbach_18, greschner_19}, quantum gases might help addressing open questions concerning the Hall effect in strongly correlated quantum phases in solid state systems~\cite{badoux_16}.
Complementarily to recent efforts in nanodevices~\cite{ella_19, bachmann_19}, they might open a new window to study ballistic magnetotransport~\cite{salerno_19, filippone_19}.
%

%
While quantum gas experiments typically focus on the measurement of the Hall polarization, the central quantity of interest in solid state systems is the Hall voltage $V_\mathrm{H}$ or the closely related Hall coefficient $R_\mathrm{H}$.
In semi-classical approaches, the latter is often interpreted as a measure of the inverse carrier density $1/\nu$~\cite{AshcroftMermin}.
For certain cases, such as noninteracting Chern-insulating states~\cite{genkina_19}, the Hall polarization $P_\mathrm{H}$ can be directly related to $V_\mathrm{H}$ or $R_\mathrm{H}$.
However, in general, this relation is nontrivial.
Thus, it is desirable to generically access $V_\mathrm{H}$ in quantum gas experiments, paving the way for a direct comparison with solid state systems.
%

%
In this Letter, for finite systems with open boundaries, we show that the Hall voltage $V_\mathrm{H}$ as well as the microscopically resolved Hall polarization $P_\mathrm{H}$ can be probed in the transient dynamics induced by suitable quantum quenches, leading to a complementary characterization of the Hall response in the interacting regime.
For the paradigmatic example of bosonic flux ladders, extensive matrix-product-state (MPS) based simulations, as well as a weak-coupling approach, reveal a remarkable robustness and zero crossings of $V_\mathrm{H}$ in different quantum phases. 
%

%
%
\begin{figure}[b]
	\includegraphics[width=\linewidth]{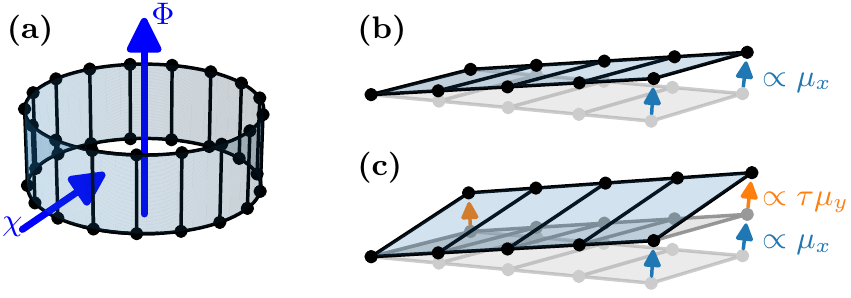}
	\caption{
		(a) Sketch of the flux-ladder ring. 
		(b) Statically tilted ladder with open boundaries.
		(c) Linear-ramp scheme for the calculation of the Hall voltage $V_\mathrm{H}$; see text.
	}
	\label{fig:sketch}
\end{figure}
{\em Hall voltage.---} We specify our approach for the case of synthetic flux ladders as realized in Refs.~\cite{atala_14, stuhl_15, mancini_15, genkina_19} and described by the Hamiltonian
\begin{align}
H = 
&- t_{x}\sum_{m=0}^{M-1}\sum_{r=0}^{L-1} 
\left(e^{i\Theta_m} a_{r,m}^\dagger a_{r+1,m} + \text{h.c.}\right) \nonumber\\
&-t_{y}\sum_{m=0}^{M-2}\sum_{r=0}^{L-1} \left(a_{r,m}^\dagger a_{r,m+1}+\text{h.c.} \right)\, + \, H_{\rm int},
\label{eq:hamiltonian}
\end{align}
with $\Theta_m=\left[m-\left(M-1\right)/2\right]\chi+\Phi/L$.
The bosonic or fermionic annihilation (creation) operator $a^{(\dagger)}_{r,m}$ acts on the $r$th rung and $m$th leg of a ladder comprising a total number of $M$ legs and $L$ rungs.
Particle hopping along the legs and rungs is parametrized by $t_x$  and $t_y$, respectively.
We typically consider site-local interactions, $H_{\rm int} = \frac{U}{2} \sum_{m,r} n_{r,m}\left( n_{r,m}-1\right)$ with $n_{r,m}=a^{\dagger}_{r,m}a_{r,m}$ and note that $\chi$ accounts for the magnetic flux piercing each plaquette.
%

%
The flux ladder Hamiltonian~\eqref{eq:hamiltonian} hosts a panoply of emergent quantum phases~\cite{orignac_01, carr_06, roux_07, tokuno_14, uchino_15, barbarino_15, petrescu_15, kolley_15, cornfeld_15, ghosh_16, greschner_16, orignac_17, strinati_17, petrescu_17, strinati_17, greschner_17}, among them Meissner phases~\cite{petrescu_13, huegel_14, piraud_15, didio_15}, with particle currents encircling the ladder along its boundaries, and vortex-lattice (VL$_{p/q}$) phases, resembling regular crystals with $p$ vortices per $(Mq)$-site unit cell~\cite{dhar_12, greschner_15}.
%

%
In ring-ladder systems with periodic boundary conditions (PBC), as shown in Fig.~\ref{fig:sketch}(a), the theoretically appealing definition of a (reactive) ground-state dc Hall response employed in Refs.~\cite{prelovsek_99, zotos_00, greschner_19, filippone_19} is based on a current-inducing Aharonov-Bohm flux $\Phi$ piercing the ring.
In general, a finite value of $\Phi$ induces a current $j_x=\frac{-i t_x}{ML}\sum_{m,r} e^{i\Theta_m} \left\langle a_{r,m}^\dagger a_{r+1,m} \right\rangle + \mathrm{h.c.}$ and a polarization $p_y = \left\langle P_y \right\rangle/(ML)$, with  $P_y=\sum_{m,r}\left[m-\left(\left(M-1\right)/2\right)\right] n_{r,m}$, giving rise to the Hall polarization $P_\mathrm{H} = p_y/j_x$.
On the other hand, the induced polarization $p_y$ might be compensated by means of an external potential term $\mu_y P_y$ in the Hamiltonian \eqref{eq:hamiltonian}, enabling the definition of the Hall voltage $V_\mathrm{H}$.
Generalizing an idea by Prelov{\v{s}}ek et al.~\cite{prelovsek_99}, in which a Hall coefficient was determined in the limit $\chi\to 0$, $V_\mathrm{H}$ is here defined for finite values of the magnetic flux $\chi$ by the requirement that $p_y$ vanishes for suitably chosen values of $\Phi$ and $\mu_y$, 
\begin{align}
V_\mathrm{H} = \mu_y / j_x.
\end{align} 
However, despite their theoretical appeal, PBC require additional engineering in typical experiments, making the systems more complex~\cite{boada_15}.
Hence, in the following, we propose alternative routes to compute the Hall voltage.
%

%
%
\begin{figure}[tb]
	\includegraphics[width=\linewidth]{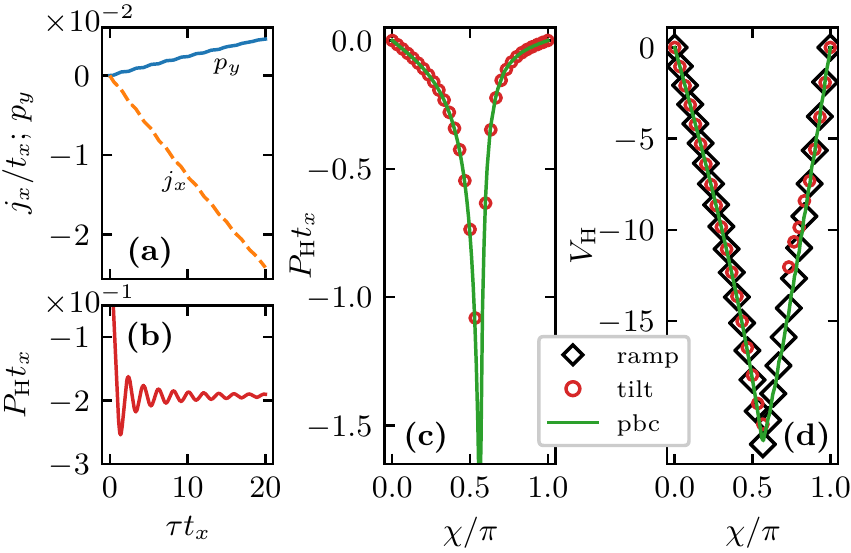}
	\caption{
		Noninteracting spinless fermions, $\nu=0.1$, $t_y/t_x = 1.6$, $M=2$. 
		(a) Transient dynamics in the current $j_x$ and in the polarization $p_y$ induced by a statically tilted potential $V_x$ with $\mu_x / t_x = 10^{-2}$ for $\chi/\pi=0.3$. 
		(b) Transient dynamics in the Hall polarization $P_\mathrm{H} = p_y/j_x$.
		(c) Hall polarization $P_\mathrm{H}$ versus magnetic flux $\chi$ as obtained from static-tilt simulations~(tilt) and adiabatic ring-ladder calculations~(pbc).
		(d) Hall voltage $V_\mathrm{H}$ versus $\chi$ as obtained from static-tilt simulations, adiabatic ring-ladder calculations, and linear potential ramps~(ramp). 
		Note that the divergence of $P_\mathrm{H}$ and the kink in $V_\mathrm{H}$ indicate the Meissner-to-vortex transition.
	}
	\label{fig:quench_free_fermions}
\end{figure}
{\em Measuring the Hall voltage.---}
In systems with open boundary conditions (OBC), the Hall voltage $V_\mathrm{H}$ can be efficiently computed within the transient dynamics induced by a \emph{linear ramp} or a \emph{static tilt}.
%

%
(i) Linear ramp. 
Starting off with the ground state of the Hamiltonian~\eqref{eq:hamiltonian}, the instantaneous turning on of a static potential $V_x = \mu_x \sum_{r,m} r n_{r,m}$ at time $\tau=0$, see Fig.~\ref{fig:sketch}(b), induces a current $j_x(\tau)$, which, in the presence of a magnetic flux $\chi$, typically polarizes the system.
However, by means of an additional time-dependent potential $V_y(\tau) = \tau \mu_y P_y$, as shown in Fig~\ref{fig:sketch}(c), the induced polarization might be compensated. 
Adjusting $\mu_y$ such that the time average of $p_y$ vanishes, $\left\langle p_y(\tau) \right\rangle_\tau= 0$, the Hall voltage can be computed as $V_\mathrm{H} =  \left\langle \mu_y \tau / j_x(\tau)\right\rangle_\tau$, where $\left\langle \bullet \right\rangle_\tau =\int_{\tau_i}^{\tau_f}\frac{\bullet}{\tau_f-\tau_i} \mathrm{d}\tau$ for a suitable time interval [$\tau_i, \tau_f$].
%

%
(ii) Static tilt.
By neglecting the dual Hall effect, referring to the current induced by the polarization itself, the Hall voltage $V_\mathrm{H}$ can be effectively calculated using a simplified protocol.
First, by instantaneously tilting the ladder by means of $V_x$, the Hall polarization $P_\mathrm{H}$ can be computed by time averaging $P_\mathrm{H} = \left\langle p_y(\tau)/j_x(\tau) \right\rangle_\tau$ in the transient dynamics.
Second, the Hall voltage $V_\mathrm{H}$ is approximated by means of $V_\mathrm{H} = P_\mathrm{H} \left( \mu_y / p_y \right)$, where $\left(\mu_y / p_y\right)$ is obtained for OBC and in the limit $\mu_y \to 0$.
%

%
The protocols are feasible in synthetic-dimension implementations~\cite{stuhl_15, mancini_15}, where the legs of the ladder correspond to different internal states of the trapped atoms.
In this case, $V_x$ and $V_y(\tau)$ can be realized by shifting the optical confining potential~\cite{genkina_19} and by detuning the internal states~\cite{lin_09a}, respectively.
Further, $j_x$ can be probed in time-of-flight measurements and Stern-Gerlach separation allows for measurements of the leg-resolved particle density, giving rise to $p_y$.
The protocols are also applicable to real-space implementations of flux ladders, in which quantum gas microscopes enable measurements of all relevant observables  and optical gradients can realize $V_x$ and $V_y(\tau)$~\cite{tai_17, zupancic_16}, as well as to continuum systems with spin-orbit coupling~\cite{lin_11, chalopin_20}.
The consistency of both protocols with the ring-ladder setup is exemplified for a noninteracting fermionic two-leg ladder in Fig.~\ref{fig:quench_free_fermions}; see below and the Supplemental Material~\cite{supmat} for further comparisons.
Figures~\ref{fig:quench_free_fermions}(a) and \ref{fig:quench_free_fermions}(b) show dynamics in $p_y$, $j_x$, and $P_\mathrm{H}=p_y/j_x$ induced by the tilt potential $V_x$.
The time-averaged results for $P_\mathrm{H}$ perfectly agree with the analytic results for PBC for $\chi\in[0,\pi]$, as shown in Fig.~\ref{fig:quench_free_fermions}(c).
The Hall voltage $V_\mathrm{H}$, shown in Fig.~\ref{fig:quench_free_fermions}(d), as well as $P_\mathrm{H}$ exhibit a nonanalyticity at the transition from a weak-flux Meissner-like region to a vortex-liquid phase~\cite{piraud_15} found for large values of $\chi$.
Moreover, as shown in Fig.~\ref{fig:quench_free_fermions}(d), $V_\mathrm{H}$ as obtained from the linear-ramp protocol perfectly agrees with the analytic results for PBC, while $V_\mathrm{H}$ as obtained from the static-tilt approximation merely deviates in the immediate proximity to the quantum phase transition.
%

%
%
\begin{figure}[b]
	\centering
	\includegraphics[width=1\linewidth]{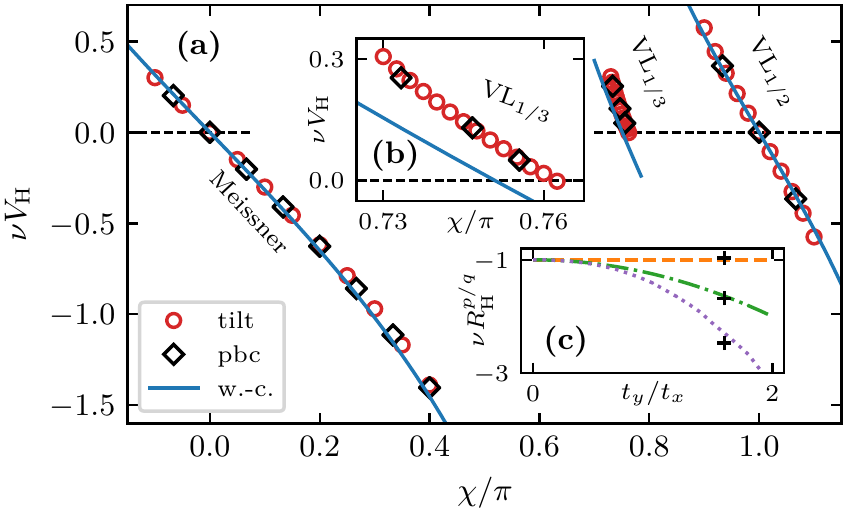}
	\caption{
		Hall voltage $\nu V_\mathrm{H}$ versus magnetic flux $\chi$ for an interacting bosonic ladder, $M=2$, $\nu=0.8$, $U/t_x=2$, $t_y/t_x=1.6$.
		(a) Symbols depict $\nu V_\mathrm{H}$ as obtained from MPS based static-tilt simulations~(tilt) and adiabatic ring-ladder calculations~(pbc) in the Meissner phase and in the vortex lattice VL$_{1/2}$ and VL$_{1/3}$.  
		The solid blue line shows the weak-coupling result (w.-c.).		
		The upper inset (b) is a close-up of the VL$_{1/3}$ data.
		The lower inset (c) shows the values of the generalized Hall coefficient $R_H^{p/q}$ for the Meissner (top dashed line) and VL phases VL$_{1/2}$ (dashed-dotted), VL$_{1/3}$  (dotted) obtained from the weak-coupling approach, showing quadratic scaling in accordance with Eq.~\eqref{eq:RHqp}; crosses depict the MPS based data. 
	}
	\label{fig:VH_chi}
\end{figure}
{\em Interacting systems.---} 
In the following, we examine the Hall voltage in bosonic flux ladders in the interacting regime.
Employing extensive MPS based simulations, performed by means of the SyTen toolkit~\cite{syten, hubigthesis_17}, we calculate the Hall voltage in quantum quenches as well as in ring-ladder setups, providing evidence for the consistency of both approaches in the strongly correlated regime.
Specifically, for ground-state calculations, we employ the single-site variant~\cite{hubig_15} of the density-matrix renormalization-group method~\cite{white_92, schollwoeck_05, schollwoeck_11}.
For quench simulations, we employ the time-dependent variational-principle algorithm~\cite{haegeman_11, paeckel_19, supmat}.
We detail on the MPS based simulations in the Supplemental Material~\cite{supmat}.
%

%
Figure~\ref{fig:VH_chi} shows the Hall voltage $V_\mathrm{H}$ for a system of strongly correlated particles ($U/t_x=2$, $t_y/t_x=1.6$) as a function of the magnetic flux $\chi$, considering an incommensurate particle filling $\nu=0.8$, where $\nu=N/(ML)$ and $N$ denotes the particle number. 
Specifically, $V_\mathrm{H}$ is shown in the Meissner  phase, in the VL$_{1/2}$ phase, and in the VL$_{1/3}$ phase~\cite{greschner_16}, noting that intermediate regions of vortex-liquid phases are omitted~\cite{vortex_liquid_remark}.
We stress that the MPS based results obtained by simulating tilt dynamics show excellent agreement with the ones obtained from ground-state calculations in ring ladders with PBC.
Moreover, our results shown in Fig.~\ref{fig:VH_chi} reveal a remarkable interaction-driven effect: a series of linear zero crossings of $V_\mathrm{H}$ in different VL phases.
%

%
In order to approach the Hall response in the VL phases from a different angle, we extend a weak-coupling (Josephson array) description~\cite{kardar_86, granato_90, mazo_95, denniston_95}, substituting in the expectation value of the Hamiltonian~\eqref{eq:hamiltonian} $a_{j,m}$ with $\sqrt{\nu_{r,m}} e^{i \theta_{r,m}}$ and introducing the classical Josephson phase $\theta_{r,m}$ and density $\nu_{r,m}$.
In the limit $t_y/t_x \to 0$ and for a homogeneous density $\nu_{r,m}=\nu$, a complete devil's staircase of such VL phases VL$_{p/q}$,  at each commensurate vortex density ${p/q}$ is predicted.
Finite values of $t_y/t_x$ and interactions gradually destabilize the VL${_{p/q}}$ phases with largest $q$~\cite{orignac_01}.
By employing the semi-classical ansatz and minimizing the energy in the presence of a current-inducing Aharonov-Bohm flux $\Phi$, we obtain the Hall voltage $V_H$ in the weak-coupling regime.
Explicitly, we find $V_\mathrm{H}=-\frac{2}{\nu}\tan\left(\chi/2\right)$ in the Meissner phase.
Moreover, in the VL$_{1/2}$ phase, we analytically find that $V_\mathrm{H}$ is independent of $U$ and proportional to $1/\nu$; see the Supplemental Material~\cite{supmat}, which details on the weak-coupling approach.
In Fig.~\ref{fig:VH_chi} the weak-coupling results are depicted by the blue solid line.
Noteworthily, they show good agreement with the MPS based results, noting that Fig.~\ref{fig:VH_chi}(b) shows deviations in the VL$_{1/3}$ phase.
%

%
Within the weak-coupling framework, the analysis of $V_H$ generically reveals a zero crossing in the center of each VL$_{p/q}$ phase at a certain value of flux $\chi_{p/q}$.
Thus, we define generalized Hall coefficients $R_H^{p/q} = \left.\partial_\chi V_H \right|_{\chi\to\chi_{p/q}}$ in analogy to the Hall coefficient obtained in the limit $\chi\to 0$~\cite{prelovsek_99}. 
Specifically, the weak-coupling approach yields
\begin{align}
R_H^{{p/q}} \approx - \frac{1}{\nu} \left[ 1 + \gamma_{p/q} \left(\frac{t_y}{t_x}\right)^2 \right]. \label{eq:RHqp}
\end{align}
We emphasize that in the Meissner phase ($\chi_0=0$) and in the VL$_{1/2}$ ($\chi_{1/2}=\pi$), Eq.~\eqref{eq:RHqp} holds exactly with $\gamma_{0}=0$, which is in accordance with Ref.~\cite{greschner_19}, and $\gamma_{1/2}=1/4$.
In the VL$_{1/3}$ phase, we find $\gamma_{1/3} \approx 0.51$ and higher order corrections in $t_y/t_x$.
The lines in Figure~\ref{fig:VH_chi}(c) depict $R_H^{p/q}$ in the Meissner, in the VL$_{1/2}$, and in the VL$_{1/3}$ phases as obtained from the weak-coupling approach.
They are in accordance with the values calculated from the MPS based data for $t_y/t_x = 1.6$.

%
%
\begin{figure}[tb]
	\centering
	\includegraphics[width=\linewidth]{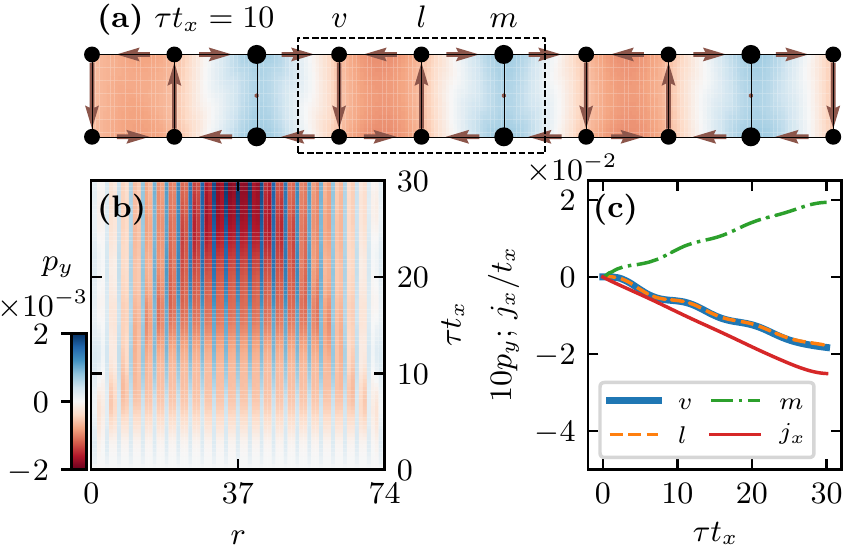}
	\caption{
		Transient dynamics induced by a static tilt in the VL$_{1/3}$, $M=2$, $\nu=0.8$, $U/t_x=2$, $t_y/t_x=1.6$, ${\mu_x / t_x = 10^{-3}}$, $\chi/\pi=0.75$, MPS simulation.
		(a) Snapshot of the ten most central rungs at time $\tau=10/t_x$ after the quench.
		The arrows show the strength of the local particle currents, the size of the dots depicts the site-local particle density, and the background shading indicates the local polarization $p_y$ of the individual rungs, using the color code from (b). 
		(b) Rung-resolved time evolution of the polarization $p_y$.
		(c) Transient dynamics in $p_y$, considering the rungs $v$, $l$, $m$ indicated in (a).
		The solid red line shows the nearly linear increase of the current $j_x$.
	}
	\label{fig:local_hall}
\end{figure}
{\em Local Hall response.---} 
Microscopic features, such as the rung-resolved polarization $p_y$, provide additional insight into the Hall response in spatially inhomogeneous VL phases.
Using MPS based simulations of the static-tilt scheme introduced above, we examine the site-resolved Hall response.
Figure~\ref{fig:local_hall}(a) depicts the local configuration of a tilted state in the VL$_{1/3}$ phase, where vortices with currents circulating counter-clockwise are surrounded by Meissner-like regions of opposite chirality.
Figures~\ref{fig:local_hall}(b) and \ref{fig:local_hall}(c) show the transient dynamics in the rung-resolved polarization.
Interestingly, the Hall response is strongly inhomogeneous, following the crystalline structure of the underlying VL phase, which remains pinned during the time evolution induced by the tilt.
In particular, we observe a positive Hall polarization of the vortices, while the Meissner-like rungs exhibit a negative Hall polarization. 
Thus, we are able to attribute to the different regions an effective local charge reflecting their Hall response: The vortices behave holelike, while the Meissner-like regions behave particlelike.
At a certain value of the magnetic flux, $\chi_{p/q}$ in each VL$_{p/q}$ phase, the competing contributions from holelike and particlelike regions cancel out, leading to a vanishing macroscopic Hall response.
The structure of the local Hall response may also be understood as a signature of the vortex-hole duality, meaning that vortices in a weakly interacting ladder may be identified with holes in a strongly interacting one-dimensional chain with a staggered potential, related to thin-torus-limit states of the fractional quantum Hall effect~\cite{greschner_17}.
%

%
Moreover, the spatially inhomogeneous Hall response following the structure of the underlying VL phases can be recovered in the weak-coupling framework, which is discussed in detail in the Supplemental Material~\cite{supmat}.
Indeed, numerical solutions confirm a direct relation between the rung-resolved polarization $p_y$ and the chirality of the local currents in the vortexlike and Meissner-like rungs, which has been tested for various VL$_{1/q}$ phases up to $q=20$.
Thus, site-resolved quantum gas microscopy~\cite{bakr_09, sherson_10, tai_17} might open a new window in the study of the Hall response, addressing local features of the Hall response and effective local charge distributions.
%

%
%
\begin{figure}[tb]
	\centering
	\includegraphics[width=1.\linewidth]{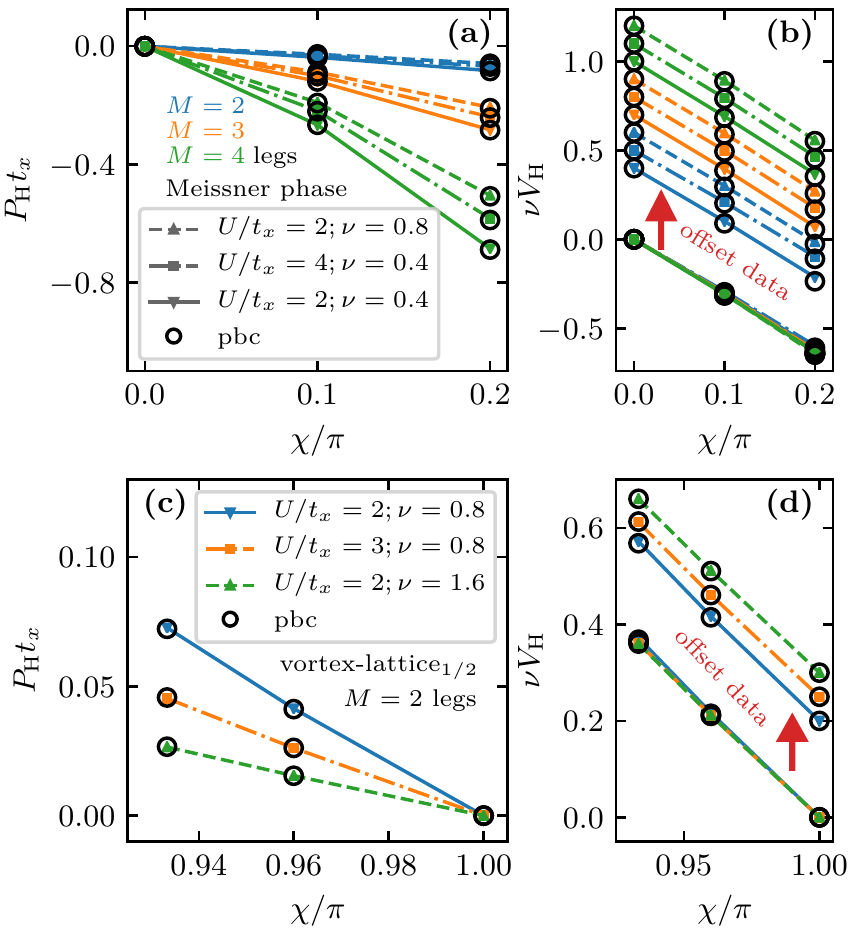}
	\caption{
		Robust Hall voltage $V_\mathrm{H}$  in the Meissner phase and in the VL$_{1/2}$, $t_y/t_x = 1.6$.
		(a)  and (b) are for the Meissner phase, showing the Hall polarization $P_\mathrm{H}$ and $V_\mathrm{H}$ as a function of the magnetic flux $\chi$ for multileg ($M=2,3,4$) ladders, different fillings $\nu=N/(LM)$, and interactions strengths $U$.
		Note that the data in (b) are vertically offset by $0.1 n+0.4$ (with $n=0,1,2,\dots$ for different values of $M$, $\nu$, and $U$) for the purpose of a clear presentation.
		(c) and (d) are for the VL$_{1/2}$ phase.
		The data in (d) are also vertically offset by $0.05 n+0.2$.
		$P_\mathrm{H}$ and $V_\mathrm{H}$ are obtained by means of static-tilt simulations (lines) and adiabatic ring-ladder calculations (open circles), as described in the text.  
		Contrarily to $P_\mathrm{H}$, the $\nu V_\mathrm{H}$ data scale on top of each other for different $\nu$, $U$, and $M$.
	}
	\label{fig:PH_VH_chi}
\end{figure}
{\em Robustness.---}
The remarkable overlap between the MPS based results for the Hall voltage $V_\mathrm{H}$ in the strongly correlated regime and the results obtained from the weak-coupling approach, as discussed in the context of Fig.~\ref{fig:VH_chi}, indicates a robustness of $V_\mathrm{H}$ with respect to the interaction strength $U$.
In Fig.~\ref{fig:PH_VH_chi} we examine this robustness in more detail, considering different values of $U$ and different particle fillings $\nu$ for various values of the magnetic flux $\chi$.
In contrast to the Hall polarization $P_\mathrm{H}$, which depends nonuniversally on the values $U$ and $\nu$, the scaled Hall voltage $\nu V_H$ collapses to one curve for a broad regime of parameters in the Meissner phase and in the VL$_{1/2}$ phase.
Moreover, in the Meissner phase, up to $M=4$ legs are considered within the adiabatic ring-ladder framework and in the static-tilt approach, revealing an additional robustness of $V_\mathrm{H}$ with respect to the ladder geometry.
For strong interactions and particle fillings close to the transition to a vortex-liquid phase, we observe deviations from the robust behavior.
We emphasize that the robustness described here is different from the universal behavior of the Hall imbalance occurring for SU(M)-symmetric interactions and small magnetic fluxes~\cite{greschner_19}, and in certain quench scenarios~\cite{filippone_19}.
%

%
%
{\em Summary.---} We have shown that the Hall voltage $V_\mathrm{H}$ can be consistently calculated in ladder systems for finite values of the magnetic flux, employing time-dependent quench protocols with longitudinal and transverse potential gradients.
The quench protocols are realistic in state-of-the-art experiments with synthetic quantum matter and a study of $V_\mathrm{H}$ in ultracold quantum gases might demonstrate its remarkable robustness with respect to the interaction strength $U$, the particle filling $\nu$, and the ladder geometry in different ground-state phases.
Furthermore, they open the exciting possibility to study $V_\mathrm{H}$ in clean and highly tunable optical lattice systems and allow for direct comparison with the Hall voltage measured in solid state devices.
A site-resolved analysis of the Hall response in vortex-lattice VL$_{p/q}$ phases provided insight into characteristic zero crossings of $V_\mathrm{H}$ at certain values of the magnetic flux $\chi_{p/q}$, where competing contributions from particle-like Meissner regions and holelike vortices cancel out.
%

%
Our schemes might prove useful in future studies of the Hall response in interesting quantum states, such as biased-ladder states~\cite{wei_14} and Laughlin-like states~\cite{cornfeld_15, petrescu_17}.
%

%
%
\begin{acknowledgments}
We thank Michele Filippone, Fabian Grusdt, and Fabian Heidrich-Meisner for inspiring discussions.
We thank Leticia Tarruell for very helpful discussions and comments.
S.G. and T.G. acknowledge support by the Swiss National Science Foundation under Division II. 
M.B. and U.S. acknowledge funding through the ExQM graduate school.
This work was supported by the Deutsche Forschungsgemeinschaft (DFG, German Research Foundation) via DFG Research Unit FOR 2414 under project number 277974659 and under Germany’s Excellence Strategy -- EXC-2111 -- 390814868.
\end{acknowledgments}
\bibliography{references}

\begin{thebibliography}{73}
\bibitem [{vor()}]{vortex_liquid_remark}%
  \BibitemOpen
  \href@noop {} {}\bibinfo {note} {Focusing on the zero crossings of
  $V_\mathrm{H}$ in the Meissner and vortex-lattice phases, we omit the
  calculation of the Hall response in vortex-liquid phases, because of its
  increased numerical complexity.}\BibitemShut {Stop}%
\end{thebibliography}
\end{document}


\title{Supplemental material to ''Probing the Hall Voltage in Synthetic Quantum Systems``}
\author{Maximilian Buser}
\affiliation{\munich} 
\author{Sebastian Greschner}
\affiliation{\geneve} 
\author{Ulrich Schollw\"ock}
\affiliation{\munich} 
\author{Thierry Giamarchi}
\affiliation{\geneve} 
\begin{abstract}
	In this supplemental material, we provide further details on the flux-ladder model, the matrix-product-state based simulations, and the weak-coupling calculations.
	We exemplify the static-tilt protocol and the linear-ramp protocol, and, moreover, we show how the spatially inhomogeneous Hall response follows the structure of underlying vortex-lattice phases in the weak-coupling framework.
\end{abstract}
\date{\today}
\maketitle
%
%
\section{Flux-ladder model}
It is recapped that in our work, we study the Hall response in the paradigmatic $M$-leg flux-ladder model, governed by the Hamiltonian
\begin{align}
H = 
&- t_{x}\sum_{m=0}^{M-1}\sum_{r=0}^{L-1} 
\left(e^{i\Theta_m} a_{r,m}^\dagger a_{r+1,m} + \text{H.c.}\right) \nonumber\\
&-t_{y}\sum_{m=0}^{M-2}\sum_{r=0}^{L-1} \left(a_{r,m}^\dagger a_{r,m+1}+\text{H.c.} \right)\nonumber\\
&+\frac{U}{2}\sum_{m=0}^{M-1}\sum_{r=0}^{L-1} n_{r,m}\left(n_{r,m}-1\right) +\mu_y P_y\,,
\label{eq:hamiltonian_sm}
\end{align}
where $\Theta_m=\left[m-\left(M-1\right)/2\right]\chi+\Phi/L$ accounts for the current-inducing Aharonov-Bohm flux $\Phi$ as well as the magnetic flux $\chi$. 
In the case of periodic boundary conditions, the latter needs to be quantized, $\chi=m2\pi/L$, with an even integer $m$ for an even number of bosons.
The bosonic or fermionic annihilation (creation) operator $a^{\left(\dagger\right)}_{r,m}$ acts on the $r$-th rung and $m$-th leg of the ladder and $n_{r,m}$ is given by $n_{r,m}=a^{\dagger}_{r,m} a_{r,m}$.
Particle hopping along the legs and rungs of the ladder is parametrized by $t_x$ and $t_y$, respectively, and $U$ quantifies the interparticle interaction strength.
We emphasize that in the Hamiltonian~\eqref{eq:hamiltonian_sm}, we consider the so-called leg gauge~\cite{greschner_16}. 
Other choices, such as the experimentally relevant rung gauge, are equivalent and do not affect the results discussed in this paper.
Moreover, in Eq.~\eqref{eq:hamiltonian_sm}, we explicitly account for the external potential $\mu_y P_y$, with
\begin{align}
	P_y = \sum_{m,r}\left(m-\frac{M-1}{2}\right) n_{r,m}\,,
\end{align}
as it plays a crucial role for the definition of the Hall voltage $V_\mathrm{H}$.
The polarization $p_y$ is defined to be an intensive quantity, given by $p_y = \left\langle P_y \right\rangle/(ML)$, noting that throughout our work, angled brackets denote expectation values.
%
%
%

%
Operators representing local particle currents can be derived from the continuity equation for the occupation of local lattice sites,
\begin{align}
\frac{d}{dt}\left\langle n_{r,m} \right\rangle =
-i \left\langle \left[n_{r,m},H\right]\right\rangle\,.
\end{align}
Thus, operators $j^\parallel_{r,m}$ and $j^\perp_{r,m}$ representing the local particle flow from site $(r,m)$ to $(r+1,m)$ and from site $(r,m)$ to $(r,m+1)$, respectively, take the form
\begin{align}
j^\parallel_{r,m}&= -i t_x e^{i\Theta(m,\chi,\phi)}a_{r,m}^\dagger a_{r+1,m}+\text{H.c.}\,,\\
j^\perp_{r,m}&=-it_ya_{r,m}^\dagger a_{r,m+1}+\text{H.c.}\,.
\end{align}
Moreover, the persistent current $j_x$ measures the unidirectional particle transport in the (ring) ladder, while the chirality $j_c(r)$ accounts for the rung-local particle flow along the outer legs in opposite directions.
They are defined by means of
\begin{align}
j_x &= \frac{1}{ML}\sum_{m=0}^{M-1}\sum_{r=0}^{L-1} j^\parallel_{r,m}\,,\\
j_c(r) &= \frac{1}{2}\sum_{r^\prime=r-1}^{r} \left( j^\parallel_{r^\prime,0}-j^\parallel_{r^\prime,M-1} \right)\,.
\end{align}
%

%
The Hall response is complementarily described by the Hall polarization $P_\mathrm{H}$ and the Hall voltage $V_\mathrm{H}$.
The former is defined in the absence of an external potential, $\mu_y=0$, 
\begin{align}
P_\text{H} = p_y/j_x\,,
\end{align}
while the latter is defined for a vanishing polarization, $p_y=0$, as realized by means of a suitably chosen $\mu_y$,   
\begin{align}
V_\text{H} = \mu_y/j_x\,.
\end{align}
%

%
Concerning open boundary conditions, the Hall polarization $P_\mathrm{H}$ and the Hall voltage $V_\mathrm{H}$ can be computed by means of the static tilt protocol and the linear-ramp protocol introduced in the main text.
While in the theoretically appealing ring-ladder setup a finite and constant current $j_x$ is induced by a finite value of the Aharonov-Bohm flux $\chi$, the time-dependent protocols are based on static tilts, which accelerate the particles in the longitudinal direction.
Hence, in the static tilt protocol, the current $\left|j_x\right|$ and the induced polarization $\left|p_y\right|$ are increasing in time.
Nonetheless, the Hall polarization $P_\mathrm{H}$ can be computed as a time ($\tau$) average $P_\mathrm{H} = \left\langle p_y(\tau)/j_x(\tau) \right\rangle_\tau$ in the transient dynamics induced by the static tilt $V_x = \mu_x \sum_{r,m} r n_{r,m}$, as described and exemplified in the main text.
Moreover, in the linear-ramp scheme an external potential $V_y(\tau) = \tau \mu_y P_y$ is ramped up in time such that the time average of the induced polarization $\left\langle p_y(\tau) \right\rangle_\tau$ vanishes.
This immediately gives rise to the Hall voltage via $V_\mathrm{H} =  \left\langle \mu_y \tau / j_x(\tau)\right\rangle_\tau$.
%

%
The time-dependent protocols can be readily realized in quantum gas experiments. 
Setups exploiting synthetic dimensions~\cite{stuhl_15, mancini_15, genkina_19} as well as setups based on quantum gas microscopes and digital micromirror devices~\cite{tai_17} are particularly well suited for this purpose. 
They provide immediate access to the current $j_x$ and the polarization $p_y$.
Moreover, the linear potentials required in the time-dependent protocols can be directly implemented in these platforms, as described in the main text.
In addition, a longitudinal particle current might also be induced by accelerating the optical lattice via a small frequency detuning of the corresponding lattice laser beams.
While the examples given in the main text are mainly focused on site-local interparticle interactions, the time-dependent protocols work equally well in the case of rung-wise interparticle interactions, which are typically realized in synthetic-dimension implementations.
Also, the relevant ground state phases which are at the focus of this paper, namely the Meissner phase and the vortex-lattice phases, can be stabilized in the presence of rung-wise interactions~\cite{greschner_16}. 
\section{Transient dynamics in the four-leg ladder}
\begin{figure}
	\includegraphics[width=\linewidth]{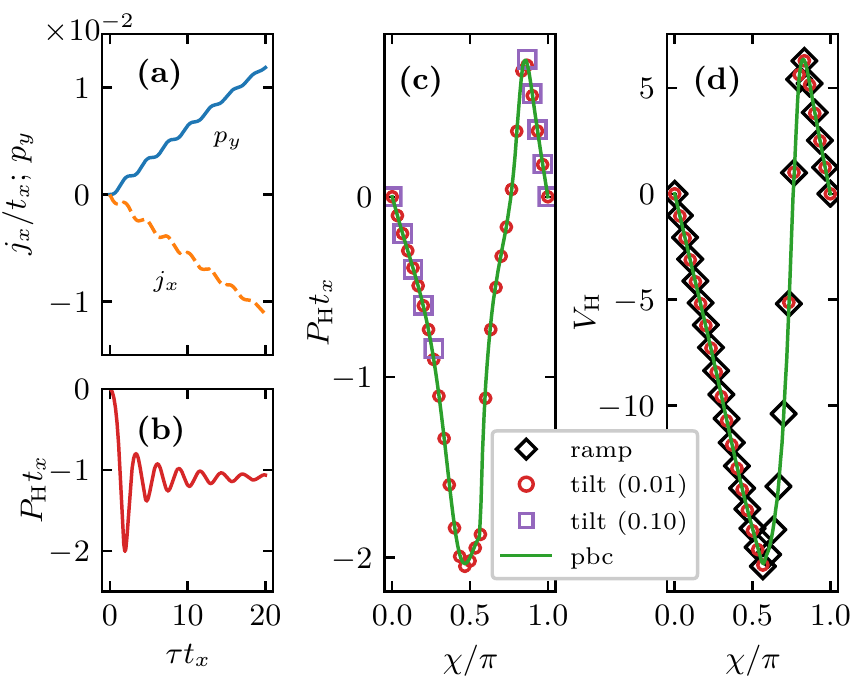}
	\caption{
		Noninteracting spinless fermions, $\nu=0.1$, $t_y/t_x = 1.6$, $M=4$. 
		(a) Transient dynamics in the persistent current $j_x$ and in the polarization $p_y$ induced by a statically tilted potential $V_x$, with $\mu_x/t_x=10^{-2}$ for $\chi/\pi=0.3$. 
		(b) Transient dynamics in the Hall polarization $P_\mathrm{H} = p_y/j_x$.
		(c) Hall polarization $P_\mathrm{H}$ versus magnetic flux $\chi$ as obtained from static tilt simulations with $\mu_x/t_x=0.01$ [tilt (0.01)], from static tilt simulations with $\mu_x/t_x=0.1$ [tilt (0.10)], and from adiabatic ring-ladder calculations~(pbc).
		(d) Hall voltage $V_\mathrm{H}$ versus $\chi$ as obtained from static tilt simulations, adiabatic ring-ladder calculations, and linear potential ramps with $\mu_x/t_x=10^{-2}$~[ramp]. 
	}
	\label{fig:quench_M4}
\end{figure}
%

%
In analogy to Fig.~2 of the main text, in Fig.~\ref{fig:quench_M4} we exemplify the different quench protocols for the case of noninteracting spinless fermions in a four-leg flux ladder.
Again, the results for the Hall polarization $P_\mathrm{H}$ and the Hall voltage $V_\mathrm{H}$ which are obtained from the linear-ramp protocol are in perfect accordance with the exact results obtained from the ground states in setups with periodic boundaries for $\chi\in[0,\pi]$.
The results obtained from the static tilt protocol with $\mu_x/t_x=0.01$ overlap well with the exact curves except for a small window of parameters in the immediate proximity to a phase transition.
The empty squares in Fig.~\ref{fig:quench_M4}(c) show that even a stronger tilt with $\mu_x / t_x=0.1$ yields the correct Hall polarization $P_\mathrm{H}$ for values of $\chi/\pi$ around zero and one.
However, the stronger tilt with $\mu_x / t_x=0.1$ is not applicable in the central region around $\chi/\pi=1/2$, where the band gap of the noninteracting model with periodic boundary conditions is reduced and the lowest band exhibits the form of a flattened out double well.
We note that apart from the parameter regime in the vicinity of the Meissner-to-vortex-liquid transition, the stronger tilt with ${\mu_x / t_x=0.1}$ is also applicable to the two-leg ladder exemplified in the context of Fig.~2 in the main text.
\section{Static tilts in the presence of interparticle interactions}
\begin{figure}
	\centering
	\includegraphics[width=1\linewidth]{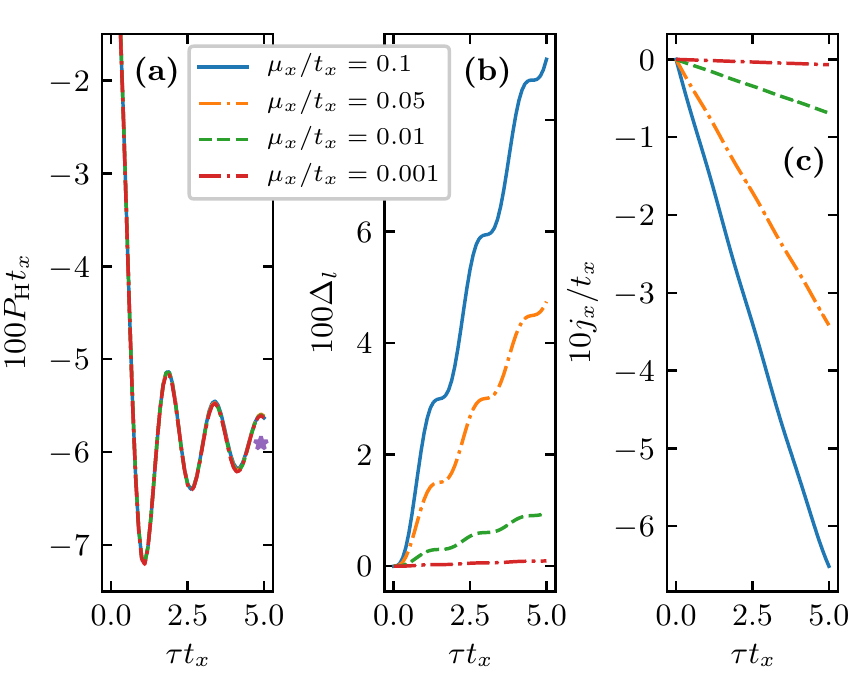}
	\caption{
		Static tilts in the interacting Meissner phase for different values of $\mu_x$, considering $\chi/\pi=0.2$, $t_y/t_x=1.6$, $U/t_x=2$, $\nu=0.8$, and $M=2$.
		(a) Transient dynamics in the Hall polarization $P_\mathrm{H}$. 
		Note that the curves for $\mu_x/t_x=0.1$, $0.05$, $0.01$, and $0.001$ fall on top of each other.
		The Hall polarization calculated in the ground state of the corresponding ring ladder is indicated by the purple star and shown for comparison.
		(b) Transient dynamics in the population imbalance $\Delta_l$ between the two legs of the ladder.
		As the particle-density profile is homogeneous in the Meissner phase, the population imbalance is the same for all rungs $r$ in the center of the system.
		(c) Transient dynamics in the current $j_x$.
	}
	\label{fig:mux}
\end{figure}
In Fig.~\ref{fig:mux}, for an interacting two-leg ladder and model parameters corresponding to the Meissner phase, we show that the static tilt protocol is applicable to a broad range of values of $\mu_x$.
Concretely, we consider $\mu_x/t_x = 0.001$, $0.01$, $0.05$, and $0.1$ for $\chi/\pi=0.2$, $t_y/t_x=1.6$, $U/t_x=2$, and $\nu=0.8$.
Figure~\ref{fig:mux}(a) shows that the transient dynamics in the Hall polarization $P_\mathrm{H}$ are independent of the considered value of $\mu_x$.
However, the current $j_x$ and the polarization $p_y$ induced by the static tilts scale roughly proportional to $\mu_x$.
Figure~\ref{fig:mux}(b) and Fig.~\ref{fig:mux}(c) show the leg-population imbalance 
\begin{align}
	\Delta_l = \left| \left\langle n_{r,0}-n_{r,1}\right\rangle\right|/\left\langle n_{r,0}+n_{r,1}\right\rangle
\end{align}
and the current $j_x$ as a function of the time $\tau$, respectively.
Note that the Meissner phase exhibits a homogeneous particle-density profile. 
Hence, the population imbalance is the same for all rungs $r$ in the center of the system.
For the strong tilt with $\mu_x/t_x=0.1$ and at time $\tau=5/t_x$ the particle numbers in the two legs differ by more than $15\%$. 
\section{Weak-coupling approach}
Putting the focus on two-leg ladders, we detail on the semiclassical weak-coupling approach~\cite{kardar_86, granato_90, mazo_95, denniston_95}.
\subsection{Semiclassical description}
Assuming a local coherent Josephson phase $\theta_{r,m}$ and a classical density $\nu_{r,m}$, we employ a coherent-state description of the ring ladder, replacing $a_{r,m}$ with $\sqrt{\nu_{r,m}} e^{i \theta_{r,m}}$ in the expectation value of the Hamiltonian~\eqref{eq:hamiltonian_sm}.
Thus, the starting point of the weak-coupling approach is
\begin{align}
\left\langle H \right\rangle =& -2t_x \sum_{r,m} \sqrt{\nu_{r,m}\nu_{r+1,m}} \cos( \theta_{r+1,m} - \theta_{r,m} + \Theta_m) \,\nonumber\\
&-2t_{y} \sum_j \sqrt{\nu_{r,m}\nu_{r,m+1}} \cos( \theta_{r,m}-\theta_{r,m+1} ) \,\nonumber\\
& + \frac{U}{2} \sum_{r,m} \nu_{r,m}(\nu_{r,m}-1) \nonumber\\
&+\mu_y  \sum_{m,r}\left(m-\frac{M-1}{2}\right) n_{r,m}\,.
\label{eq:H_JJ}
\end{align}
This ansatz, which generically addresses the regime of weak but finite interaction strengths $U$ (similar to Josephson junction arrays), as typically given in experiments with large bosonic particle fillings $\nu$, is well suited for the description of vortex-lattice (VL$_{p/q}$) phases. 
\begin{figure}
	\centering
	\includegraphics[width=1\linewidth]{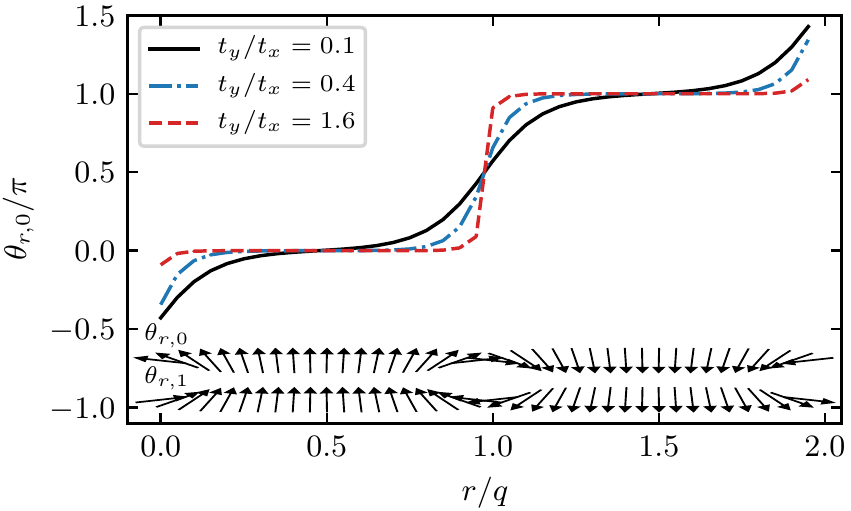}
	\caption{
		Josephson phase $\theta_{r,m}$ in the weak-coupling approach for the VL$_{1/20}$ phase considering $U/t_x=2$, $\nu=0.8$, $\chi/\pi=0.2$, and different values of $t_y/t_x$. 
		The lower arrows sketch the real space behavior of the phase in the $m=0$ and $m=1$ leg of the ladder for $t_y/t_x=0.1$. 
		Vortices, corresponding to the $\pi$-phase slips, delocalize as $t_y/t_x$ decreases.
	}
	\label{fig:classVL_psi}
\end{figure}
Typical low-energy configurations of the Josephson phase $\theta_{r,m}$ are shown in Fig.~\ref{fig:classVL_psi}.
They exhibit a regular series of localized vortices where $\theta_{r,m}$ slips by $\pi$, while in the intermediate regions the phases $\theta_{r,0}$ and $\theta_{r,1}$ are aligned, similar to a small Meissner phases.
Moreover, the vortices delocalize as $t_y/t_x$ decreases and typical low-energy configurations satisfy $\left(\theta_{r+1,0}-\theta_{r,0}\right)\approx-\left(\theta_{r+1,1}-\theta_{r,1}\right)$.
%

%
For the practical calculation of the Hall response in the weak-coupling approach, as well as for the calculation of the configurations shown in Fig.~\ref{fig:classVL_psi}, we generically consider a homogeneous particle density per rung, employing the following parametrization
\begin{align}
	\nu_{r,0} &= 2\nu \cos^2\left(\alpha_r\right)&~~\nu_{r,1} &= 2\nu \sin^2\left(\alpha_r\right)\,,
\end{align}
and minimize $\left\langle H \right\rangle$, as given in Eq.~\eqref{eq:H_JJ}, with respect to the parameters $\alpha_{r}$ and $\theta_{r,m}$, noting that $r\in\left\lbrace 0,1,\dots,L-1 \right\rbrace$ and $m\in\left\lbrace 0,1 \right\rbrace$.
Concerning the Hall voltage $V_\mathrm{H}$, $\mu_y$ is considered as a Lagrange multiplier, while the Hall polarization $P_\mathrm{H}$ is obtained for $\mu_y=0$. 
Thus, in the Meissner phase, which can also be understood as the VL$_0$ phase, we find
\begin{align}
	V_\mathrm{H} = -\frac{2}{\nu}\tan\left(\frac{\chi}{2}\right)\,,
\end{align}
as stated in the main text.
Moreover, the weak-coupling result for $V_\mathrm{H}$ in the VL$_{1/2}$ phase, as discussed in the context of Fig.~3 of the main text, is explicitly given by
\begin{align}
V_\mathrm{H}=
-\frac{1}{\nu}\frac{2 \sin (\chi )
	\left(\frac{t_y^2}{t_x^2}-2 \cos (\chi
	)+2\right)}{\left(\frac{t_y^2}{t_x^2}+4\right) \cos (\chi
	)+\frac{t_y^2}{t_x^2}-\cos (2 \chi )-3}\,.
\end{align}
Concerning VL$_{1/q}$ phases with $q>2$, we numerically find (local) minima of $\left\langle H \right\rangle$ in the vicinity to an initial configuration given by $\alpha_{r}=\pi/4$ and $\theta_{r,m}=\pi\left\lfloor r/q \right\rfloor$ for $r\in\left\lbrace 0,1,\dots,L-1 \right\rbrace$ and $m\in\left\lbrace 0, 1\right\rbrace$.
Thus, the VL$_{p/q}$ configurations shown in Fig.~\ref{fig:classVL_psi} do not necessarily correspond to the true ground state in the weak-coupling model~\eqref{eq:H_JJ} (for the considered model parameters) but resemble metastable configurations for a fixed vortex filling with $p=1$ vortices in $q=20$ rungs.
%

%
%
%
\subsection{Rung-resolved Hall polarization}
In accordance with our matrix-product-state based results addressing the strongly interacting regime, the rung-resolved Hall polarization $P_\mathrm{H}$, as obtained from the weak-coupling approach, follows the local structure, and, specifically the chirality $j_c(r)$, of the underlying vortex-lattice phases.
\begin{figure}[tb]
	\centering
	\includegraphics[width=1\linewidth]{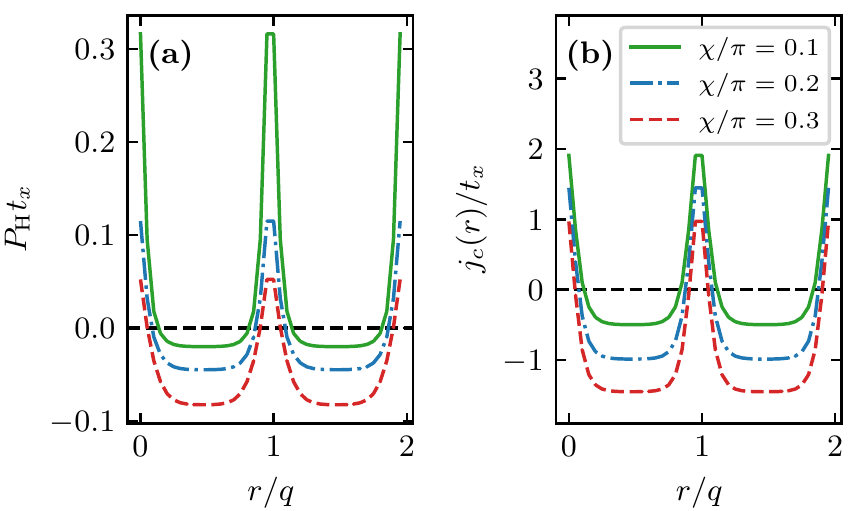}
	\caption{
		(a) Rung-resolved Hall polarization $P_\mathrm{H}$ and (b) chirality $j_c(r)$ in the VL$_{1/20}$ phase, considering $t_y/t_x=0.4$, $U/t_x=2$, $\nu=0.8$, and different values of the magnetic flux $\chi$.
		Note that for the model parameters considered here, the VL$_{1/20}$ configurations are metastable solutions, which do not necessarily correspond to the ground state in the weak-coupling model.
	}
\label{fig:classVL_py}
\end{figure}
Figure~\ref{fig:classVL_py}(a) and Fig.~\ref{fig:classVL_py}(b) show $P_\mathrm{H}$ and $j_c(r)$ in the VL$_{1/20}$ phase for different values of the magnetic flux $\chi$.
The vortices exhibit a holelike Hall response with $V_\mathrm{H}>0$ and $j_c>0$, while in the surrounding particlelike Meissner regions one finds $V_\mathrm{H}<0$ and  $j_c<0$.
\section{Matrix-product-state based calculations}
Here, we provide additional information on the matrix-product-state based approaches employed for the calculation of the Hall response in the strongly correlated regime, acknowledging that both, ground-state calculations as well as time-dependent calculations, are performed by means of the SyTen toolkit~\cite{syten, hubigthesis_17}.
Throughout our work, the $U(1)$ symmetry of the flux-ladder Hamiltonian~\eqref{eq:hamiltonian_sm} corresponding to the conservation of the total particle number is enforced on the level of the matrix-product-state tensors.
Moreover, a cutoff to at most six bosons per lattice site is sufficient for the model parameters considered in our work.
\subsection{Ground states in the ring-ladder setup}
The ground states in the ring-ladder setup are calculated by means of the density-matrix renormalization-group method~\cite{white_92,schollwoeck_05,schollwoeck_11}.
Specifically, we employ the single-site variant of the algorithm using subspace expansion~\cite{hubig_15}. 
In the course of these calculations, we consider bond dimensions up to typically $3000$.
%

%
In general, periodic boundary conditions complicate the variational ground-state optimization, and, in practice, they require an increased amount of sweeping, as compared to the optimization in analogous systems with open boundary conditions.
Moreover, the quantization of the magnetic flux $\chi$, as discussed in the context of the Hamiltonian~\eqref{eq:hamiltonian_sm}, manifests itself as a challenging constraint. 
Especially for ground-state phases appearing in a narrow window of $\chi$, large systems may need to be considered.
Specifically, in order to resolve the VL$_{1/3}$ phase in Fig.~3 of the main text, we consider ladders with $L=60,~75,~\textrm{and}~90$ rungs in the ring-shaped setup.
For the calculation of ground states in the presence of a finite current-inducing Aharonov-Bohm flux $\Phi$, we generically employ the ground states attained at $\Phi=0$ as an initial state.
%

%
Convergence of the variationally optimized ground states is ensured by means of a comparison of the energies $\left\langle H \right\rangle$, the energetic fluctuations $\left\langle H^2 \right\rangle-\left\langle H \right\rangle^2$, as well as all relevant global and local observables for different bond dimensions and different values of the site-local bosonic cutoff.
Additionally, it is ensured that the ground states in the Meissner phase and in the vortex-lattice phases exhibit regular patterns of well-defined unit cells.
\subsection{Time-dependent simulations}
The static tilt protocol introduced in the main text is simulated using the two-site variant of the time-dependent variational-principle algorithm~\cite{haegeman_11, paeckel_19} after obtaining the ground state in a ladder with open boundaries from a preliminary density-matrix renormalization-group calculation as described above.
In the presence of interactions, we verify the time-averaged results for the Hall polarization $P_\mathrm{H}$ by considering various values of the tilt parameter $\mu_x$ typically ranging from $\mu_x/t_x=10^{-4}$ up to $\mu_x/t_x=10^{-1}$.
For the propagation in time, we employ bond dimensions up to $500$ and ensure convergence in all relevant observables by varying the time-step size and the maximum bond dimension independently.
Finally, the consistent Hall response, which is independently obtained from either time-dependent quench simulations or ground-state calculations in ring-ladder setups, confirms our results and the feasibility of both approaches. 
\bibliography{references}